\documentclass[letter,twocolumn]{jpsj2} 

\usepackage{graphicx}
\usepackage{dcolumn}
\usepackage{bm}

\title{Photogenerated Carriers in SrTiO$_3$ Probed by 
Mid-Infrared Absorption}

\author{%
Hidekazu \textsc{Okamura}\thanks{E-mail: okamura@kobe-u.ac.jp}, 
Masato \textsc{Matsubara}, 
Koichiro \textsc{Tanaka}$^1$, 
Kazutoshi \textsc{Fukui}$^2$, 
Mitsushi \textsc{Terakami}$^3$, 
Hideyuki \textsc{Nakagawa}$^3$, 
Yuka \textsc{Ikemoto}$^4$, 
Taro \textsc{Moriwaki}$^4$, 
Hiroaki \textsc{Kimura}$^4$ and 
Takao \textsc{Nanba}
}

\inst{%
Graduate School of Science and Technology, Kobe University, 
Kobe 657-8501 
\\
$^1$Department of Physics, Graduate School 
of Science, Kyoto University, Kyoto 606-8502. 
\\
$^2$Research Center for Development of Far-Infrared Region, 
Fukui University, Fukui 910-8507.
\\
$^3$Department of Electrical and Electronics 
Engineering, Fukui University, Fukui 910-8507.
\\
$^4$Japan Synchrotron Radiation Research 
Institute and SPring-8, Sayo, Hyogo 679-5198.
}

\recdate{\today}

\abst{
Infrared absorption spectra of SrTiO$_3$ have been measured 
under above-band-gap photoexcitations to study the properties 
of photogenerated carriers, which should play important roles 
in previously reported photoinduced phenomena in SrTiO$_3$.     
A broad absorption band appears over the entire mid-infrared 
region under photoexcitation.      
Detailed energy, temperature, and excitation power dependences 
of the photoinduced absorption are reported.  
This photo-induced absorption is attributed to the intragap 
excitations of the photogenerated carriers.     
The data show the existence of a high density of in-gap states 
for the photocarriers, which extends over a wide energy range 
starting from the conduction and valence band edges.    
}

\kword{SrTiO$_3$ (strontium titanate), quantum paraelectric, 
photocarriers, infrared spectroscopy}

\begin{document}
\maketitle

Structural, electronic, and magnetic changes in solids 
induced by photoexcitations have attracted much interest 
recently, since a photoexcitation technique allows one 
to control these properties in unique manners that cannot 
be achieved by varying other external parameters such as 
temperature, pressure, electric and magnetic 
fields.\cite{nasu}     
One of such photoinduced phenomena is an enhancement of 
dielectric properties under photoexcitation in a range 
of dielectrics called ``quantum paraelectrics'', a most 
well-known example of which is SrTiO$_3$ (STO).

STO is an insulator with a band gap of about 3.2~eV.   
Upon a photoexcitation above the band gap energy, 
it gives a broad photoluminescence (PL) band between 
1.6 and 3.0 eV.\cite{PL-1,PL-2}   
The Stokes shift between the absorption edge 
and the center of PL band is about 0.8~eV, 
showing that the photogenerated electron-hole pairs 
experience large energy relaxations before recombination.      
Time-resolved PL\cite{PL-3,Tanaka-PL1} and spectral hole 
burning\cite{Tanaka-PL2} experiments of STO have shown 
that the photogenerated carriers (photocarriers) reach a 
localized state before recombination, which has been 
discussed in terms of self-trapped excitons, polarons 
and strong electron-lattice 
coupling.\cite{PL-2,PL-3,Tanaka-PL1,Tanaka-PL2}  
Figure~1 illustrates the situation in STO expected 
from these optical results.     
STO is also well known for a high 
photoconductivity.\cite{PL-2,photocon}    
According to photo-Hall studies, the majority photocurrent 
carriers are negatively charged, with a mobility as high as 
10$^4$~cm$^2$/Vsec at 5~K.\cite{photoHall-1,photoHall-2}     
Regarding the photocarriers in STO, it has been debated 
how their high mobility shown by transport experiments 
and their localized characters shown by optical experiments 
can be understood in a consistent manner.\cite{nasu2}

STO has a large static dielectric constant of 
$\epsilon_1 \sim$ 200 at 300~K, which further increases upon 
cooling, reaching $\sim$ 20000 at 4~K.   In spite of the 
extremely large $\epsilon_1$, STO remains paraelectric down 
to 0.3~K due to a quantum paraelectricity.\cite{muller}      
Recently, it has been found that the static dielectric 
constant of STO remarkably increases under photoexcitations 
above the band gap energy.\cite{takesada,tanaka}    
It is likely that the photocarriers also play an important 
role in this phenomenon, but its microscopic mechanism is 
not clear yet.

In this work, we study the infrared absorption in STO 
under photoexcitation, to probe the properties of 
photocarriers.       
A broad absorption band has been observed over the entire 
mid-infrared under a photoexcitation at 3.4~eV.    
This photo-induced infrared absorption is attributed to 
intragap excitations of carriers by the infrared photons, 
rather than to their Drude response.      
The present results demonstrate that there are a 
high density of in-gap states for the photocarriers, 
and that the in-gap states are distributed over a 
wide energy range reaching the band edges.

The single crystal sample of STO, purchased from 
Earth Chemical Co., Ltd., was initially a plate of 0.5~mm 
thickness, with polished (110) faces.    To avoid detecting 
internally reflected light beam, one of the two faces 
was further polished to create a wedge of about 
3~$^\circ$ between them.      
The IR transmission of STO under photoexcitation were 
measured at the beam line BL43IR of SPring-8,\cite{okamura} 
using a Fourier-transform infrared interferometer and a 
HgCdTe detector.      
The UV photoexcitation was provided by a frequency-doubled, 
mode-locked Ti:sapphire laser operating at 84.7~MHz.    
The photon energy (wavelength) of the laser was set 
to 3.4~eV (365~nm), which was well above the band gap energy 
(3.2~eV) of STO.        
Special care was taken so that the IR beam arising from 
the reflection off the rear surface of the sample was 
rejected, 
and that the excitation laser beam covered the entire 
(2~mm diameter) area of the sample through which 
the IR beam transmitted.    
The PL of the same sample was measured using 
the same laser source.   
Furthermore, the reflectivity spectrum $R(\omega)$ of the same 
sample was measured at photon energies 0.01 - 30~eV 
(without photoexcitation).    The dielectric function 
$\epsilon(\omega)$ and the absorption coefficient 
$\alpha(\omega)$ were obtained by Kramers-Kronig (K-K) 
analysis of the measured $R(\omega)$.\cite{wooten}    
The experimental details of the $R(\omega)$ measurements were 
similar to those reported elsewhere.\cite{YbAl3}

Figure~2(a) shows the infrared transmittivity spectrum 
$T(\omega)$ of STO at 8~K without photoexcitation.    
Here, $T(\omega)$ is defined simply as the transmitted IR spectrum 
with the sample divided by that without the sample.      
The strong absorption 
below $\sim$ 0.15~eV is due to optical phonons.   The dips 
observed at 0.22 and 0.27~eV were previously discussed in 
terms of defect-related absorption.\cite{defect}          
When a UV photoexcitation was made on the sample, $T(\omega)$ was 
observed to {\it decrease} slightly, i.e., {\it a photoexcitation 
induced an additional IR absorption} in STO.     
Figure~2(b) plots the measured photoinduced IR absorption 
(PIA) spectrum $A(\omega)$ at various 
temperautres.    Here, $A(\omega)$ has been defined as 
\begin{equation}
A(\omega)=-\frac{T_{\rm on}(\omega)-T_{\rm off}(\omega)}{T_{\rm off}}=
1-\frac{T_{\rm on}(\omega)}{T_{\rm off}(\omega)}, 
\end{equation}
where $T_{\rm on}(\omega)$ and $T_{\rm off}(\omega)$ are 
$T(\omega)$ spectra measured with the excitation laser 
on and off, respectively.   
It is seen that a broad absorption band has been induced by 
the photoexcitation over a wide energy range.       
Although not shown here, the absorption persisted up to 
2~eV (the high-energy limit of our measurement), in agreement 
with a previous report.\cite{Tanaka-PL2}     
In addition to a broad component rising toward lower energy, 
a shoulder is observed at 0.23~eV [indicated by the arrow in 
Fig.~1(b)].     This shoulder becomes clear only below 100~K.  
When the laser was turned off, the absorption disappeared 
immediately.     The temperature below which the PIA appeared, 
about 130~K, well coincided with that below which the PL was 
strong enough to be observed by the naked eye.  
Figure~2(c) shows $A(\omega)$ at 8~K recorded under different 
laser powers ($P_{\rm laser}$).  
     The appearance and growth of PIA with 
increasing $P_{\rm laser}$ is very similar to that with 
decreasing temperature in Fig.~1(b).      
Figure~2(d) compares $A(\omega)$ at 8~K with the 
temperature-induced absorption without photoexcitation, 
1-$T$(20~K)/$T$(8~K).    
The photo- and temperature-induced absorption spectra 
are apparently very different; i.e., the PIA is not due 
to a sample heating by the laser.  
Similar measurements were also made in a reflection 
geometry.  The $R(\omega)$ spectra of the same (wedged) 
sample showed almost negligible photoinduced change.    
However, $R(\omega)$ of another STO sample {\it without a wedge} 
exhibited a photoinduced decrease (not shown here) very 
similar to $A(\omega)$ in Fig.~2.   In this case $R(\omega)$ 
contained the IR beam reflected from the rear face of 
the sample (hence going through the sample).     
These results demonstrate that the observed PIA results 
from the photocarriers {\it inside} the sample, 
rather than a change in $R(\omega)$ at the surface.

We have measured detailed temperature and $P_{\rm laser}$ 
dependences of PIA.        To quantitatively analyze the data, 
we integrate $A(\omega)$ between 0.18 and 0.55~eV, and use 
the integrated intensity as a measure of the PIA.          
The time-averaged intensity of PL was also recorded using 
the same laser source with the same (3.4~eV) photon energy.  
(A Si photodiode detector was used without a spectrometer, 
so that the total PL intensity in the 1.1-3.2~eV 
range was recorded.)     
Figure~3 summarizes the results, where the intensities 
have been normalized to the maximum value in each graph.    
The PL intensity is almost constant 
below 35~K, but it decreases rapidly above 35~K, in 
agreement with previous report.\cite{Tanaka-PL2}      
Then the PL intensity shows a plateau at 70-90~K range.  
In contrast, the PIA decreases more gradually with 
increasing temperature, with two plateaus at 30-50~K and 
70-90~K ranges.      Above 70~K, the PL intensity is only 
$\sim$ 5\% of the maximum intensity, while the PIA is still 
30\% of the maximum.     Note that both PIA and PL have a 
plateau at almost the same temperature range of 70-90~K.         
Regarding the $P_{\rm laser}$ dependence, the PL shows almost 
linear dependence, although the slope is slightly larger 
at low-power region below 0.2~W/cm$^2$.    On the other 
hand, the PIA increases much more rapidly with 
$P_{\rm laser}$ below 0.1~W/cm$^2$ than above.    
Namely, the dependences of PIA and PL on the temperature 
and $P_{\rm laser}$ share some common features, 
but they are quantitatively very different.

Now we shall consider possible mechanisms by which the 
photocarriers give rise to the observed PIA.  
Since STO shows high photoconductivity with a mobility reaching 
10$^4$~cm$^2$/Vs,\cite{photoHall-1,photoHall-2} 
an apparent possibility is a free-carrier 
absorption.     Below, we use the Drude model to calculate 
the absorption coefficient $\alpha_{\rm D}(\omega)$ due to 
free carriers having a density $n$ and an effective mass $m^\ast$.   
The real ($\epsilon_1$) and imaginary ($\epsilon_2$) dielectric 
functions are expressed as:\cite{wooten}  
\begin{equation}
\epsilon_1(\omega) = 
\epsilon_\infty - \frac{\omega_p^2}{\omega^2 + \tau^{-2}}, \hspace{0.25cm} 
\epsilon_2(\omega) = \frac{\tau^{-1}}{\omega} \cdot
\frac{\omega_p^2}{\omega^2 + \tau^{-2}}
\end{equation}
where $\epsilon_\infty$ is the contribution from higher-energy 
interband transitions, $\omega_p = 4\pi n e^2/m^\ast$ the plasma 
energy, $\tau$ the average relaxation time of the carriers, and 
$\omega$ the photon energy.     $\epsilon_\infty$ = 6.9 was obtained 
from the $\epsilon_1(\omega)$ given by the K-K analysis of 
the measured $R(\omega)$.    $\tau$ was 
estimated from the previously reported value of 
$\mu$=10$^4$~cm$^2$/Vs at 5~K,\cite{photoHall-2} assuming the free 
carrier relation $\mu = e \tau / m^\ast$.    Then 
$\epsilon_1$ and $\epsilon_2$ can be calculated through 
Eq.~(2) for given values of $m^\ast$ and $n$.   
The imaginary refractive index is given as 
$\kappa = \frac{1}{\sqrt{2}}[-\epsilon_1 + 
\sqrt{\epsilon_1^2 + \epsilon_2^2}]^{1/2}$, and the 
absorption coefficient due to free carriers is finally 
obtained as 
$\alpha_{\rm D} = (4\pi /c)\omega \kappa$.\cite{wooten}    
In a photo-Hall 
experiment,\cite{photoHall-2}  $n \sim$ 5 $\times$ 
10$^{14}$~cm$^{-3}$ has been obtained under a photon flux 
density of 2 $\times$ 10$^{16}$~cm$^{-2}$sec$^{-1}$, 
assuming that the photocarriers are uniformly distributed 
within a plate of thickness $t$=10~$\mu$m.    
Our maximum photon flux density is $\sim$ 5~$\times$ 
10$^{18}$~cm$^{-2}$sec$^{-1}$, hence the corresponding $n$ 
could be $\sim$ 10$^2$ times larger than the above value 
for the same value of $t$.    
The penetration depth at the laser photon energy (3.4~eV) 
was $\sim$ 0.2~$\mu$m from the absorption coefficient of 
STO at 3.4~eV ($\simeq$ 5~$\times$ 10$^4$~cm$^{-1}$), 
obtained from the K-K analysis of measured $R(\omega)$.     
Since the photocarriers have a high mobility, $t$ may 
be larger than the penetration depth of the laser, but 
$t$=10~$\mu$m might be an overestimation.    
If a value of $t$=1~$\mu$m is used instead, $n$ becomes 
10 times larger than that in Ref.~9.    
Taking into account these effects of $P_{\rm laser}$ and 
$t$, we estimate $n$ to be 10$^{17}-10^{18}$~cm$^{-3}$ 
for the maximum laser power 
in this study.       
Figure~4(a) shows $\alpha_{\rm D}(\omega)$ calculated for 
$m^\ast = m_0$ and $n$=10$^{17}$, 10$^{18}$, and 
10$^{19}$~cm$^{-3}$.       
Regarding $m^\ast$, no data are available for photocarriers 
in STO, but $m^\ast = 5-20~m_0$ have been reported for 
chemically carrier-doped STO.\cite{mass}     
Using these values instead of $m_0$ would further reduce 
$\alpha_{\rm D}$ compared with those in Fig.~4(a).

Next, we will estimate the absorption coefficient due to the 
photocarriers, $\alpha_{\rm ph}(\omega)$, using the 
present PIA data.      We again assume that the PIA 
occurs within a plate of thickness $t$.   Then one may 
express the experimental data in terms of $\alpha_{\rm ph}$ as: 
\begin{equation}
T_{\rm on} = T_{\rm off} \exp(-\alpha_{\rm ph}t)
\end{equation}
Figure~4(b) plots $\alpha_{\rm ph}(\omega)$ obtained 
using Eq.~(3) and the $A(\omega)$ spectrum at 8~K under 
$P_{\rm laser}$=1.3~W/cm$^2$ for $t$=1~$\mu$m.          
[$\alpha_{\rm ph}$ for $t$=10~$\mu$m is, for example, 
simply 0.1 times that in Fig.~4(b).]   
Comparing Figs. 4(a) and 4(b), it is clear that the 
free-carrier absorption for the values of $n$ and $t$ 
expected for the current study is too small to account 
for the observed PIA.

Having seen that a Drude response of photocarriers cannot 
account for the PIA, the most likely mechanism for the PIA 
should be excitations of photocarriers to above (below) 
the conduction (valence) band edge, as schematically 
shown by the process (4) in Fig.~1.     
In this case, apparently the intensities of PL 
($I_{\rm PL}$) and PIA ($I_{\rm PIA}$) should be 
proportional to the time-averaged photocarrier density 
$\bar{n}$.    
However, there should be other factors affecting 
$I_{\rm PL}$ and $I_{\rm PIA}$, since they have different 
dependences on the temperature and $P_{\rm laser}$ as 
observed in Fig.~3.    Suppose that a laser pulse 
initially creates $n_0$ photocarriers, which subsequently 
decay with a mean lifetime of $\tau$.\cite{Tanaka-PL2}    
Here, $\tau$ includes both radiative ($\tau_{\rm r}$) and 
non-radiative ($\tau_{\rm nr}$) decays, 
$\tau^{-1}=\tau_{\rm r}^{-1} + \tau_{\rm nr}^{-1}$.   
Then $\bar{n} = (1/T) \int_0^T n_0 e^{-t/\tau} dt = 
n_0 \tau / T$, where $T$ is the pulse interval, and one obtains 
the following relations:     
\begin{eqnarray}
I_{\rm PL} & = & (1/\tau_{\rm r}) \bar{n} \propto n_0 \tau/\tau_{\rm r}, \\
I_{\rm PIA}& = & (1/\tau_{\rm ir}) \bar{n} \propto n_0 \tau/\tau_{\rm ir}. 
\end{eqnarray}
Here 1/$\tau_{\rm r}$ and 1/$\tau_{\rm ir}$ are the probabilities 
(oscillator strengths) of the radiative (PL) recombination and 
the IR absorption, respectively.     Hence, the different 
temperature and $P_{\rm laser}$ dependences of $I_{\rm PL}$ and 
$I_{\rm PIA}$, observed in Fig.~3, are probably due to different 
dependences of $\tau_{\rm r}$ and $\tau_{\rm ir}$ on these 
parameters.       
Below 35~K, both the time-averaged intensity and the mean 
lifetime of the PL was found almost constant,\cite{Tanaka-PL2} 
hence $\bar{n}$ in STO should be independent of temperature 
below 35~K.    Then the observed increase of $I_{\rm PIA}$ with 
decreasing temperature even below 35~K should be due to a change in 
$1/\tau_{\rm ir}$.    
On the other hand, the common feature in $I_{\rm PL}$ and 
$I_{\rm PIA}$ in Fig.~3, i.e., the presence of a plateau 
at 70-90~K, probably results from the temperature dependence 
of $\bar{n}$, which is commonly contained in both $I_{\rm PL}$ 
and $I_{\rm PIA}$.

The present PIA data show two important results: 
(i) The density of in-gap states occupied by the 
photocarriers is high, as shown by the large magnitude 
of $\alpha_{\rm ph}$ [Fig.~4(b)].       
(ii) The distribution of the in-gap states extends 
continuously to the conduction and valence band edges, 
as shown by the increasing PIA toward $\omega$=0.    
The situation is illustrated in Fig.~5.   
Of course, the observation of strong PL well below the absorption 
edge had previously shown a presence of in-gap 
states, and its broad bandwidth had also shown thier distribution 
over a wide energy range.\cite{PL-1,PL-2,PL-3}     
However, it is well known in semiconductor physics\cite{cardona} 
that even a very low density of impurities or defects may give 
rise to a strong PL below the band gap energy.    In such a case, 
the impurities or defects would show only a weak or negligible 
absorption, in striking contrast to the present observation of 
large $\alpha_{\rm ph}$ in SrTiO$_3$.       
It is also well known that the PL recombination tends to occur 
at a lower-energy part of the DOS spectrum, so that an observed 
PL spectral shape does not necessarily show the true spectral 
shape of the PL-emitting DOS.       
Since $A(\omega)$ increases with decreasing $\omega$ down to 
the lowest measured energy, it is likely to keep increasing 
as $\omega$ approaches zero.   Namely, the in-gap DOS 
should extend to the band edges as shown by the thin solid 
curves in Fig.~5, in contrast to what would be suggested 
by the PL spectrum alone (dotted curves).

The present results do not explicitly show whether or not the 
photocarriers giving rise to the PIA are localized.    
However, previous results of 
time-resolved PL experiments\cite{PL-3,Tanaka-PL1,Tanaka-PL2} 
strongly suggest that, after a 
photoexcitation, the photocarriers quickly relax in energy 
to localized in-gap states, from which the PL recombinations 
take place.     As shown by the present study, the density 
of these in-gap states is quite high.   Therefore, it is very 
likely that the 
photocarriers are mobile only for a short time period right 
after their generation, and that they occupy the in-gap, 
localized states for the rest of their lifetime.   
This is equivalent to stating that only a small fraction 
of the photocarriers are mobile at a given time.     
To actually study how the photocarriers relax in energy 
as a function of time, it would be necessary to perform a 
time-revolved PIA experiment.    
Note also that one cannot distinguish contributions by the 
electrons from those by the holes in the present PIA study.   
Although the spectra in Fig.~2 show {\it two} distinct spectral 
components, namely a broad component rising toward lower 
energy and a narrower one peaked at 0.23~eV, it is 
unclear whether or not they correspond to electrons 
and holes.      
A time-resolved PIA experiment might reveal different 
temporal dynamics corresponding to electrons and holes, 
since the conduction electrons in STO, having a strong Ti 
3$d$ character,\cite{band} are expected to experience 
a stronger electron correlation and a stronger 
electron-phonon coupling than the holes having an 
O 2$p$ character.

In conclusion, SrTiO$_3$ has shown a broad photoinduced 
absorption band over the entire mid-IR region 
under an above-band-gap photoexcitation.   Their detailed 
energy, temperature, and excitation power dependences 
have been reported.     The photoinduced absorption 
has been interpreted as arising from excitations of 
photocarriers to the conduction and valence bands.   
The present result shows that there is a high density 
of in-gap states available for the photocarriers, 
and that they are distributed over a wide energy range 
extending to the conduction and valence band edges.      
In considering the properties of photocarriers in SrTiO$_3$, 
these results should be taken into account in addition 
to the previous photoluminescence results.    
A time-resolved photoinduced IR absorption experiment 
is proposed as a future study to further clarify the 
temporal relaxation of photocarriers.      

\section*{Acknowledgements} 
We thank Prof. K. Nasu for useful discussions.    
The experiment at SPring-8 was performed under the 
approval of JASRI (2003A0433-NS1-np and 2004A0779-NSa-np).   
We thank Dr. T. Hirono for technical assistance in the 
PL measurement.  
A part of the reflectivity measurement was performed as 
the Joint Studies Program of the Institute for Molecular 
Science (2003).

\pagebreak

%
\begin{figure}
\begin{center}
\caption{Processes in SrTiO$_3$ discussed in the text.   
(1) Laser photons ($h \nu_{\rm laser}$) excite electrons from 
the valence band (VB) to the conduction band (CB).   
(2) The photocarriers relax non-radiatively to in-gap 
states via the electron-lattice coupling.   
(3) When an electron-hole pair recombine radiatively, 
the photoluminescence results ($h \nu_{\rm PL}$).    
(4) When an electron (a hole) absorbs an infrared photon 
($h \nu_{\rm IR}$) before recombination, it is excited back 
to CB (VB).  
}
\end{center}
\end{figure}

\begin{figure}
\begin{center}
\caption{(a) Transmittivity [$T(\omega)$] of SrTiO$_3$ at 7~K.   
(b) Photoinduced absorption $A(\omega)$ under an excitation 
laser power ($P_{\rm laser}$) of 1.3~W/cm$^2$ at 
different temperatures, and (c) that at 8~K under 
different $P_{\rm laser}$'s.     The vertical arrows 
indicate the shoulder discussed in the text.   
(d) Temperature-induced and photoinduced absorption 
spectra at 8~K.   For the former, the difference 
between 20 and 8~K is shown.   
}
\end{center}
\end{figure}
%
\begin{figure}[t]
\begin{center}
\caption{
(Left graph): Photoinduced absorption (PIA) of SrTiO$_3$ 
integrated over 0.18-0.55~eV range and the photoluminescence 
(PL) intensity as a function of temperature under 
two laser powers ($P_{\rm laser}$).      
(Right graph): PIA and PL of STO as a function of 
$P_{\rm laser}$ at 8 and 80~K.    
In these graphs, the intensities have been normalized by 
the maximum values.    
}
\end{center}
\end{figure} 

\begin{figure}[t]
\begin{center}
\caption{
(a) Theoretically-calculated absorption coefficient due to 
free carriers ($\alpha_{\rm D}$) having the effective 
mass $m^\ast = m_0$ for three different values of 
carrier density ($n$).      
(b) Experimentally-obtained absorption coefficient due 
to photocarriers ($\alpha_{\rm ph}$), derived from the 
PIA data at 8~K under $P_{\rm laser}$=1.3~W/cm$^2$ 
in Fig.~1 for $t$=1~$\mu$m.      
}
\end{center}
\end{figure} 

\begin{figure}[t]
\begin{center}
\caption{
Illustration of the density of states (DOS) as a function of 
energy (E) in SrTiO$_3$.      
The thick solid curves indicate the band edges for conduction 
band (CB) and the valence band (VB).   The thin solid curves 
indicate the in-gap DOS for the photocarriers, suggested by 
the present PIA data.    
The dotted curves indicate an in-gap DOS implied by the PL 
spectral shape alone.    
The horizontal arrows represent the relevant optical 
transitions, and for the purpose of illustration the 
situation is assumed to be symmetric about the middle of 
the band gap.     
}
\end{center}
\end{figure} 

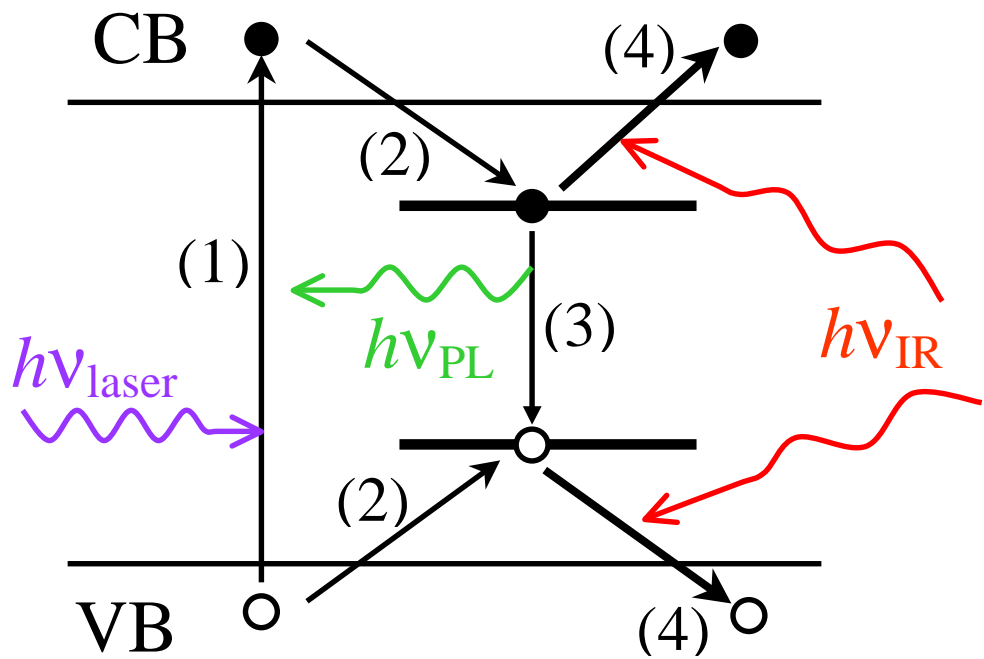
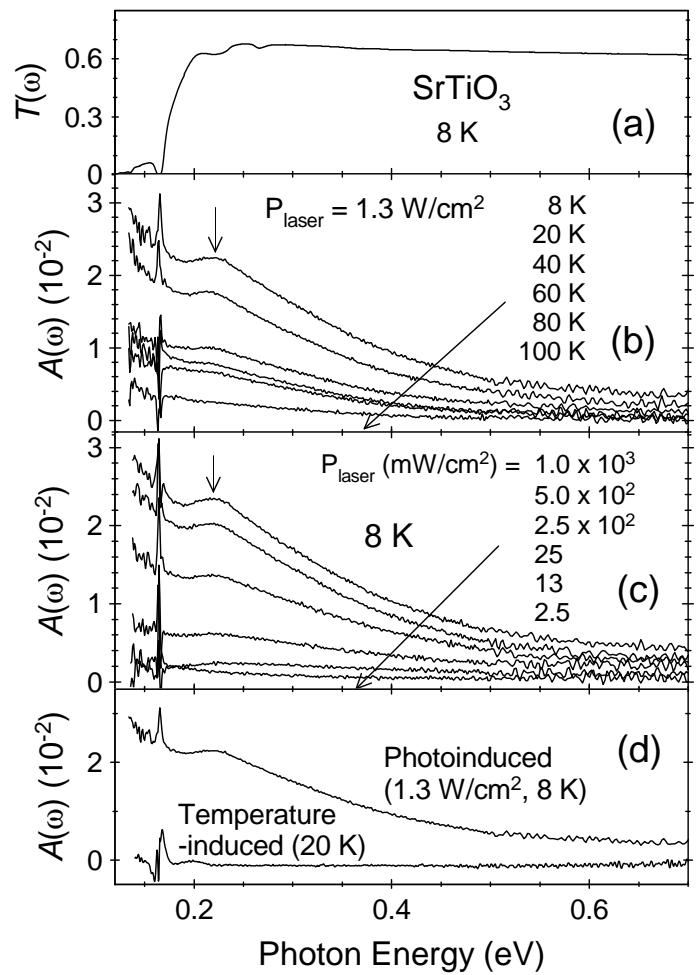
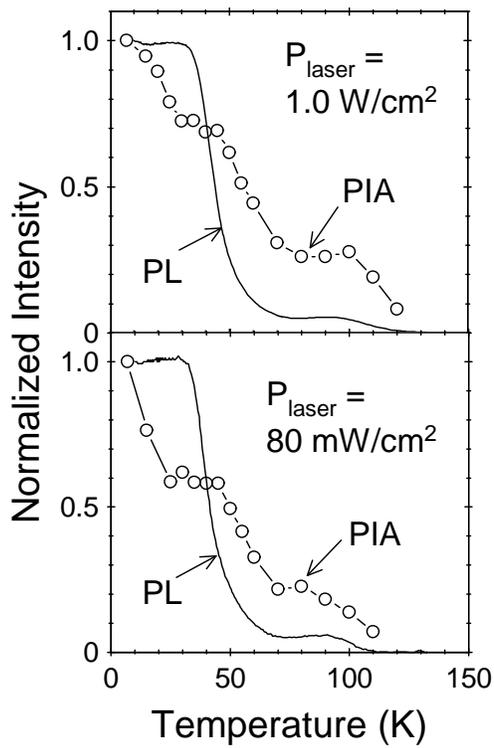
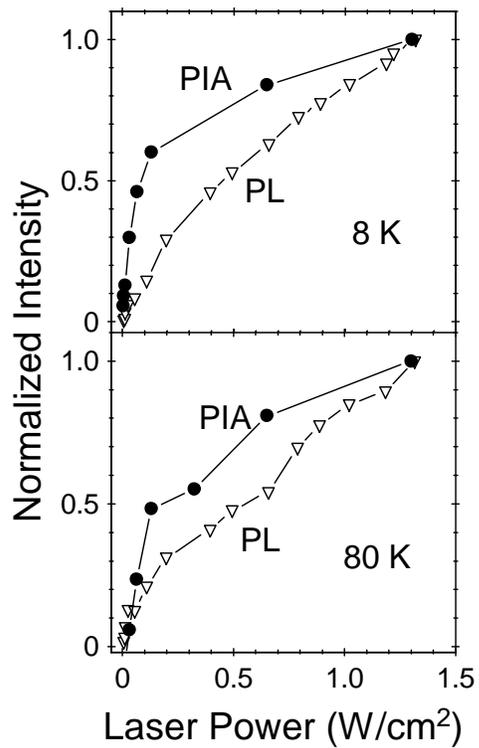
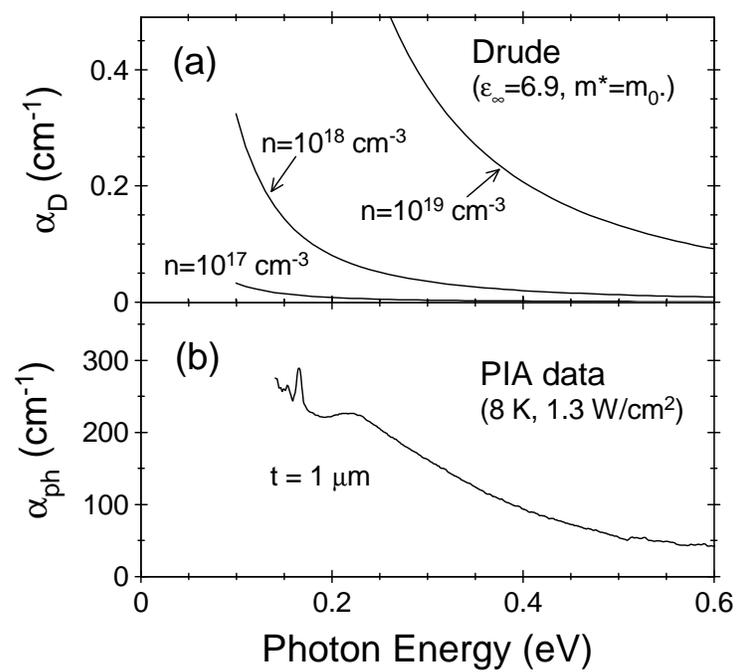
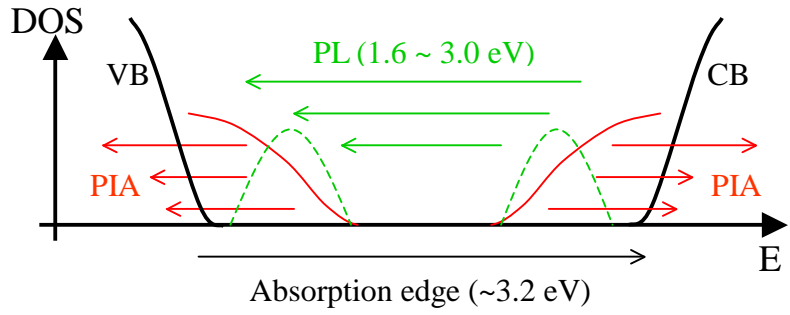
\end{document}